\let\texprimitiveyear\year
\documentclass[doublecol]{epl2}

\usepackage{amsmath,amssymb}
\usepackage{graphicx}
\usepackage{booktabs}
\usepackage[table]{xcolor}
\let\eplyearcommand\year
\let\year\texprimitiveyear
\usepackage{tikz}
\usetikzlibrary{positioning,arrows.meta}
\let\year\eplyearcommand

\newcommand{\Psync}{P_{\mathrm{sync}}}

\title{Arithmetic of the sync basin for pulse-coupled oscillators}
\shorttitle{Arithmetic of the synchronization basin}

\author{K. P. O'Keeffe}
\shortauthor{K. P. O'Keeffe}

\institute{\inst{1} Starling Research Institute --- Seattle, WA 98112, USA}


\abstract{A population of $N$ identical pulse-coupled oscillators ultimately
settles into one of two outcomes: full synchrony or a state of co-existing
synchronized clusters. We show that which outcome occurs is controlled by the
prime factorization of $N$. At the critical charging curve --- linear, the
boundary between the synchronizing and clustering regimes --- the
synchronization basin acquires exact arithmetic structure. The synchronization
probability is $\Psync=A_{N,1}/N^N$ for all $N$, where $A_{N,1}$ satisfies
an exact recurrence relation. For prime $N$, $A_{N,1}=N^N-1$ giving the closed
form $\Psync=1-1/N^N$; for composite $N$, the observed asymptotic scaling is
$1-\Psync\sim C_m N^{-(m-1)}$, where $m$ is the smallest prime divisor.
The result adds a new member to the atlas of exotic basin
geometries: alongside fractal, riddled, and tentacled basins, we now have a
basin that is arithmetic.}

\begin{document}
\maketitle

\section{Introduction}
Pulse-coupled oscillators (PCOs) are a canonical model of collective
synchronization in biological and engineered
systems~\cite{Peskin1975,Mirollo1990,Winfree1980,Pikovsky2001,Strogatz2000}.
Each unit charges toward a threshold and, on firing, delivers a pulse that
advances its neighbors. For the standard concave (leaky) charging curve,
Mirollo and Strogatz proved the celebrated result that an all-to-all population
synchronizes from almost all initial conditions~\cite{Mirollo1990}, the route
proceeding through clusters of synchronized oscillators that gradually coalesce~\cite{Kirst2009,OKeeffe2015,OKeeffe2016,Timme2002,Kaneko1990}.

Most work on PCOs asks \emph{whether} or \emph{how fast} synchrony occurs; the complementary question
--- how large is the synchronization basin? --- is harder and exact answers are
rare~\cite{Wiley2006,Menck2013}. We answer it for the \emph{convex} charging
regime ($\gamma<0$ in the standard $\dot x = 1-\gamma x$ model), which removes
the Mirollo--Strogatz guarantee: synchrony is no longer assured, and persistent
cluster states become possible. Convex charging is not a contrived case: it occurs
in quadratic and exponential integrate-and-fire neuron
models~\cite{Ermentrout1996,FourcaudTrocme2003,Badel2008} and captures the
accelerating spike-initiation dynamics of many excitable cells. Knowing the size
of the sync basin may have practical value: the heart, for instance, has two
competing attractors --- normal rhythm and ventricular fibrillation --- and the
basin of the former determines how robustly the heart resists sudden cardiac
death~\cite{Wiley2006}. Our results give the first exact basin-volume
calculations in the idealized, PCO setting.

Basin geometry has surprised before: fractal boundaries~\cite{McDonald1985},
riddled basins~\cite{Alexander1992,Ott1994}, and ``octopus'' basins whose volume
hides in thin tentacles~\cite{ZhangStrogatz2021} each revealed that the set of
initial conditions leading to a given attractor can be far stranger than
intuition suggests. Here we find a basin phenomenon of a different kind: not shape,
but arithmetic. For linear charging curves $\gamma=0$, we find the exact result
$\Psync(N,0)=A_{N,1}/N^N$ for general $N$ with $A_{N,1}$ satisfying a recurrence relation.
When $N$ is prime, $A_{N,1}=N^N-1$ giving $\Psync = 1-1/N^N$. For composite $N$, we compute the first few $A_{N,1}$ exactly, and find empirically that the asymptotic scaling is $1-\Psync \sim C_m N^{-(m-1)}$, where $m=\mathrm{lpf}(N)$, the least prime factor of $N$. Table~\ref{tab:results} summarizes these core findings.
\begin{table}[t]
\centering
\caption{Main results. At $\gamma=0$ the synchronization basin is determined exactly for all $N$; for $\gamma<0$ exact results are limited to $N\le5$. $m=\mathrm{lpf}(N)$ is the least prime factor; $A_{N,1}$ is computed by an exact recurrence relation (SM~\S2); the composite scaling law (exponent and prefactors) is empirical (SM~\S4).}
\begin{tabular}{lll}
\toprule
\rowcolor{blue!20}
Setting & $\Psync$ & Status \\
\midrule
\rowcolor{blue!8}
$\gamma=0$, $N$ prime      & $1-1/N^N$                         & exact \\
\rowcolor{blue!8}
$\gamma=0$, $N$ composite  & $A_{N,1}/N^N$                     & exact \\
\rowcolor{blue!8}
$\gamma=0$, $N$ composite  & $1-C_m N^{-(m-1)}$               & empirical \\
\rowcolor{orange!12}
$\gamma\in[-1,0]$, $N\le5$ & eqs.~(\ref{eq:p2})--(\ref{eq:p5}) & exact \\
\bottomrule
\end{tabular}
\label{tab:results}
\end{table}

\section{Model}
We take $N$ identical oscillators with voltage
$x_i\in[0,1]$ obeying
\begin{equation}
  \dot x_i = 1-\gamma x_i
  \label{eq:model}
\end{equation}
on a complete graph. The charging curve $x(t)=(1-e^{-\gamma t})/\gamma$ is
concave for $\gamma>0$, convex for $\gamma<0$, and linear at $\gamma=0$.
When $x_i=1$ the unit fires, resets to $0$, and delivers a pulse $\Delta=k/N$
to all others, where $k$ is its cluster size; any neighbor pushed to
$x_j+\Delta\ge1$ merges into the firing cluster. Initial voltages are
i.i.d.\ Uniform$(0,1)$, and $\Psync(N,\gamma)$ denotes the probability of
reaching full synchrony. We study $\gamma\le0$.

\section{Linear charging curves $\gamma=0$}
The synchronization basin is the set of initial conditions $x\in[0,1]^N$ from
which the system reaches full synchrony; its volume is $\Psync(N)$. In general
this is hard to compute: $\Psync$ is the volume of a region whose boundaries
are implicitly defined by the nonlinear firing dynamics, and tracking which
initial conditions lead to sync requires following every possible sequence of
firing and absorption events.

The $\gamma=0$ limit is however tractable. At linear charging, voltage is
phase: $x_i(t)=x_i(0)+t$ between firings. This means the Poincar\'e return
map after one full firing round is a pure translation --- cluster $j$ shifts
by $p_0-p_j$, where $p_i=k_i/N$ is its pulse fraction. For a trajectory to
avoid absorption forever, this shift must vanish for every cluster, forcing
$p_j=p_0$ for all $j$. This is the \emph{equal-size principle}: every
non-synchronizing trajectory settles into equal-size clusters whose common
size divides $N$, with $K\mid N$ clusters each of size $N/K$ (SM~\S2).
Moreover every divisor is realized --- for each $K\mid N$ an explicit
construction witnesses a $K$-cluster terminal state --- so the number of
distinct outcomes is exactly $d(N)$, the number of divisors of $N$ (SM~\S2).
The arithmetic thinning is drastic: a priori the population could end in any
of Euler's $p(N)\sim e^{\pi\sqrt{2N/3}}/(4N\sqrt3)$ partitions; the dynamics
retains only the $d(N)$ equal-divisor ones --- $9$ rather than
$1.9\times10^8$ at $N=100$.

To see the principle in action, consider $N=4$ with two clusters of sizes
$(3,1)$. Their pulse fractions are $p_0=3/4$ and $p_1=1/4$, so the return
map shifts the gap by $3/4-1/4=1/2$ per round. The gap grows without bound
and the smaller cluster is eventually absorbed. By contrast, two clusters of
sizes $(2,2)$ have $p_0=p_1=1/2$: the gap is frozen and the state persists
forever. Equal sizes are the only survivors. For a larger example --- $N=15$, a product of two primes --- see SM~\S4.

\begin{figure}[t!]
  \centering
  \includegraphics[width=\columnwidth]{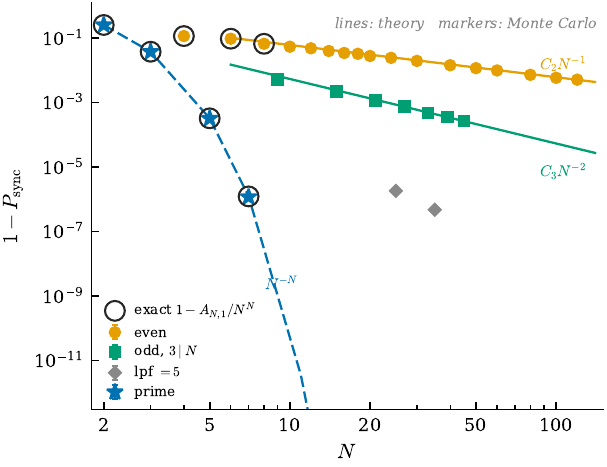}
  \caption{Synchronization basin at $\gamma=0$: theory against measurement.
  Markers are Monte Carlo; lines and rings are theory, with no fitted parameter
  except the two prefactors $C_2\approx0.60$ and $C_3\approx0.53$. Open rings:
  the exact enumeration $1-A_{N,1}/N^N$ of eq.~(\ref{eq:word}), available for
  $N\le8$. Dashed: the exact prime law $N^{-N}$, eq.~(\ref{eq:main}). Solid: the
  composite asymptotics $C_mN^{-(m-1)}$ of eq.~(\ref{eq:scaling}), with
  $m=\mathrm{lpf}(N)$.}
  \label{fig:comp}
\end{figure}

For prime $N$ the only non-sync outcome is the all-singleton state $(1)^N$.
Its probability is $1/N^N$ --- the volume of the polytope where all $N-1$
ordered gaps exceed $1/N$ --- giving our first main result,
\begin{equation}
  \Psync(N) = 1 - \frac{1}{N^N}, \qquad N \text{ prime},
  \label{eq:main}
\end{equation}
which agrees with Monte Carlo simulations across all primes tested
(fig.~\ref{fig:comp}).

For general $N$, write each initial voltage as $Nx_i=a_i+r_i$ with
$a_i\in\{0,\ldots,N-1\}$ integer and $r_i\in[0,1)$ fractional. At $\gamma=0$
firing and absorption depend only on which oscillator is ahead of which ---
that is, on the rank order of the fractional parts $r_i$, not their actual
values. Now \emph{relabel the oscillators in increasing order of $r_i$} and call
the resulting integer string $a=(a_0,\ldots,a_{N-1})$ the \emph{word}. By
exchangeability the word is uniform on $\{0,\ldots,N-1\}^N$ and independent of
the ordered fractional parts, and in this labelling the dynamics is purely
integral: writing $d=b_j-b_f+s_f$ for a non-firing cluster $j$ and a firer $f$
of mass $s_f$, cluster $j$ is absorbed iff $d>0$, or $d=0$ and $q_j>q_f$, where
$b$ is the integer phase and $q$ the rank of a cluster's leader (SM~\S2). The
outcome is therefore constant on each of the $N^N$ equally likely words, and the
continuous basin-volume problem reduces to counting how many words lead to
sync. Letting $A_{N,1}$ be that count,
\begin{equation}
  \Psync(N) = \frac{A_{N,1}}{N^N},
  \label{eq:word}
\end{equation}
where $A_{N,1}$ is computed by a finite recurrence on cluster states
(SM~\S2). Explicitly, a state $S=((b_0,s_0),\dots,(b_{L-1},s_{L-1}))$ lists
the $L$ clusters' integer phases and masses in rank order. The next firer is
$f=\arg\max_i(b_i,i)$; it absorbs every $j$ allowed by the rule above, and
each survivor moves to $b_j'=b_j-b_f+N+s_f$, defining a map $S\mapsto T_NS$.
Since $L$ never increases and the state space is finite, iterating $T_N$
either reaches a single cluster or revisits a state. Setting $\sigma(S)=1$ in
the first case and $0$ in the second, so that $\sigma(S)=\sigma(T_NS)$,
\begin{equation*}
  A_{N,1}=\sum_{a\in\{0,\dots,N-1\}^N}\sigma\bigl((a_0,1),\dots,(a_{N-1},1)\bigr).
\end{equation*}
Exact enumeration gives
\begin{equation}
  \frac{A_{N,1}}{N^N}:\quad
  \frac{3}{4},\ \frac{26}{27},\ \frac{227}{256},\ \frac{3124}{3125},\
  \frac{42275}{46656},\ \frac{823542}{823543},\ \frac{15682639}{16777216}
  \label{eq:exact_vals}
\end{equation}
for $N=2,\dots,8$. For prime $N$, $A_{N,1}=N^N-1$ consistent
with eq.~(\ref{eq:main}), as the entries at $N=2,3,5,7$ show. Figure~\ref{fig:comp} shows these exact values
agree with Monte Carlo across all $N$ tested.

The structure is sharpest for $N=pq$, a product of two distinct primes $p<q$.
Then $d(N)=4$ and the four outcomes --- one per divisor --- are synchrony
$(N)^1$, $p$ clusters of size $q$, $q$ clusters of size $p$, and all singletons
$(1)^N$, with weights
\begin{equation*}
  1-C_p N^{-(p-1)},\quad C_p N^{-(p-1)},\quad C_q N^{-(q-1)},\quad N^{-N}.
\end{equation*}
For $N=15$ ($p=3$, $q=5$) simulation gives synchrony at probability $0.997852$,
the $(5)^3$ state at $2.14\times10^{-3}$, the $(3)^5$ state at $7.7\times10^{-6}$,
and $(1)^{15}$ at $15^{-15}=2.28\times10^{-18}$ (SM~\S4).

How does $1-\Psync(N)$ scale as $N\to\infty$? For prime $N$ eq.~(\ref{eq:main})
gives $N^{-N}$, which is negligible. The interesting case is composite $N$,
where each proper divisor opens an $m$-cluster sector; the dominant one is
set by the smallest prime divisor $m=\mathrm{lpf}(N)$ (fig.~\ref{fig:comp}):
\begin{equation}
  1-\Psync(N)\ \sim\ C_m\,N^{-(m-1)},
  \label{eq:scaling}
\end{equation}
where both the exponent and the prefactors $C_2\approx0.60$,
$C_3\approx0.53$ are empirical: the equal-size principle dictates which
sectors occur but not their weights, and a proof of
eq.~(\ref{eq:scaling}) --- a local-limit property of the entrance into $m$
equal blocks --- remains open (SM~\S4).

Together, eqs.~(\ref{eq:main})---(\ref{eq:scaling}) give a complete picture
of $\Psync(N)$ at $\gamma=0$: exact for all $N$ via eq.~(\ref{eq:word}), with
a transparent asymptotic structure controlled by the arithmetic of $N$.

\section{Convex charging, $\gamma<0$}

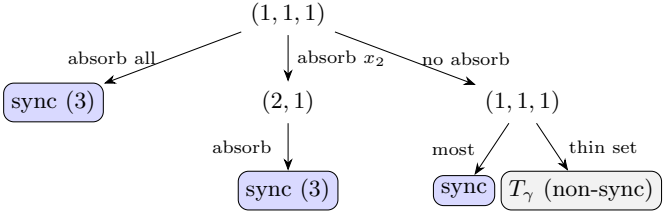
\begin{figure}[t!]
\centering
\resizebox{\columnwidth}{!}{%
\begin{tikzpicture}[
  every node/.style={font=\small},
  level 1/.style={sibling distance=32mm, level distance=12mm},
  level 2/.style={sibling distance=16mm, level distance=12mm},
  edge from parent/.style={draw, -Stealth, thin},
  term/.style={draw, rounded corners, inner sep=3pt, fill=gray!10},
  sync/.style={draw, rounded corners, inner sep=3pt, fill=blue!15},
]
\node {$(1,1,1)$}
  child { node[sync] {sync $(3)$}
          edge from parent node[left, xshift=-1mm] {\scriptsize absorb all} }
  child { node {$(2,1)$}
          child { node[sync] {sync $(3)$}
                  edge from parent node[left, xshift=-1mm] {\scriptsize absorb} }
          edge from parent node[right, xshift=0mm] {\scriptsize absorb $x_2$} }
  child { node {$(1,1,1)$}
          child { node[sync] {sync}
                  edge from parent node[left, xshift=-1mm] {\scriptsize most} }
          child { node[term] {$T_\gamma$ (non-sync)}
                  edge from parent node[right, xshift=1mm] {\scriptsize thin set} }
          edge from parent node[right, xshift=1mm] {\scriptsize no absorb} };
\end{tikzpicture}
}%
\caption{Branch decomposition for $N=3$. Starting from three singletons
$(1,1,1)$, the first firing (oscillator $3$) branches into three cases.
Blue: synchrony. Grey: the thin invariant set $T_\gamma$ whose volume gives
$1-\Psync(3,\gamma)$ exactly.}
\label{fig:branch}
\end{figure}

For $\gamma<0$ the equal-size principle no longer holds: unequal clusters can
persist, and the basin must be computed directly. A formula for general $N$
remains out of reach, but for $N\le5$ we find one by tracing all possible
cluster trajectories --- a branch decomposition.

Consider $N=2$ oscillators. Order the voltages $x_1\le x_2$; oscillator $2$
fires first and delivers a pulse $1/2$ to oscillator $1$. If the pulse is large
enough to push oscillator $1$ to threshold, they merge immediately and sync.
If not, oscillator $1$ advances but misses, and a simple return-map argument
shows it never catches up. So $\Psync(2,\gamma)$ is the probability of
first-event absorption,
\begin{equation}
\begin{split}
  \Psync(2,\gamma) = 2\int_0^1\int_0^{x_2}
  \mathbf{1}\!\left[x_2 \le \frac{1+2(1-\gamma)x_1}{2-\gamma}\right]
  dx_1\,dx_2,
\end{split}
  \label{eq:int2}
\end{equation}
where the indicator is $1$ when oscillator $2$'s pulse pushes oscillator $1$
to threshold (absorption), and $0$ otherwise. This evaluates to eq.~(\ref{eq:p2}).

\begin{figure}
  \centering
  \includegraphics[width=\columnwidth]{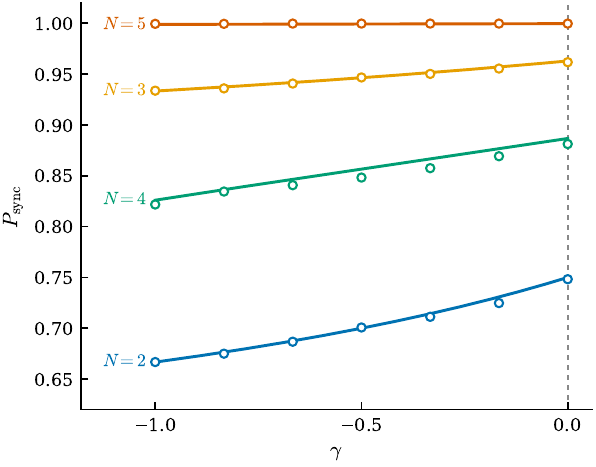}
  \caption{Exact $\Psync(\gamma)$ for $N=2,3,4,5$ (curves, eqs.~(\ref{eq:p2})--(\ref{eq:p5}))
  versus Monte Carlo (markers). All jump at the critical point $\gamma=0$ (dashed).}
  \label{fig:gamma}
\end{figure}

$N=3$ follows the same logic but with more branches. Order
$x_1\le x_2\le x_3$; oscillator $3$ fires first. Its pulse either absorbs
oscillator $2$ (forming a 2-cluster and a singleton), absorbs neither, or syncs all
three immediately. In the 2-cluster-plus-singleton branch a return-map inequality shows the
pair always absorbs the singleton. In the no-absorption branch most initial
conditions eventually absorb, but a thin invariant set $T_\gamma$ never does
--- its volume, a rational integral in closed form, is exactly
$1-\Psync(3,\gamma)$. Figure~\ref{fig:branch} illustrates the tree for $N=3$.
The same architecture extends to $N=4,5$ with more
branches but identical logic (SM~\S1). This yields

\begin{align}
  \Psync(2,\gamma) &= \frac{3-\gamma}{2(2-\gamma)},\label{eq:p2}\\
  \Psync(3,\gamma) &= 1-\frac{(1-\gamma)^2}{3(3-\gamma)(3-2\gamma)},\label{eq:p3}\\
  \Psync(4,\gamma) &= 1-P_{2+2}(\gamma)-\frac{(1-\gamma)^3}{8(4-\gamma)(2-\gamma)(4-3\gamma)},\label{eq:p4}\\
  \Psync(5,\gamma) &= 1-\frac{(1-\gamma)^4}{5(5-\gamma)(5-2\gamma)(5-3\gamma)(5-4\gamma)},\label{eq:p5}
\end{align}
on $\gamma\in[-1,0]$, where $P_{2+2}(\gamma)$ is the contribution from the
$2{+}2$-cluster channel of $N=4$; it is an exact rational function, too lengthy
to display, with $P_{2+2}(0)=7/64$ (SM~\S1). All four agree with Monte
Carlo to statistical error (fig.~\ref{fig:gamma}). Each is strictly below $1$
for $\gamma<0$ and jumps at $\gamma=0$.

Each of eqs.~(\ref{eq:p2})--(\ref{eq:p5}) can be extended below $\gamma=-1$,
but each breaks down at a sharp endpoint $\gamma_\ast(N)$ below which new
unequal-cluster attractors appear and the formula no longer applies; the
endpoints and their mechanisms are derived in SM~\S1. Together,
eqs.~(\ref{eq:p2})--(\ref{eq:p5}) give exact basin volumes for all
$\gamma\in[-1,0]$ and $N\le5$; all jump discontinuously at $\gamma=0$
(fig.~\ref{fig:gamma}).

\section{Discussion}
We have found exact basin volumes at $\gamma=0$ for all $N$, and exact formulas
for $N\le5$ across $\gamma\in[-1,0]$. The central finding is that the
synchronization probability develops exact number-theoretic structure at the
critical point $\gamma=0$: the basin is governed by the prime factorization of
$N$, and where fractal or riddled basins are remarkable in their geometry, this
one is remarkable in its arithmetic.

The arithmetic is however \emph{fragile} because the translation law used in
the equal-size proof is special to linear charging. For $\gamma<0$, the taming
coordinate $w=1-\gamma x$ turns free charging into a common rescaling and a
pulse into an additive shift, so a full return is M\"obius (more generally,
projective), not a size-only translation.

For two clusters of sizes $n_1\ge n_2$, the exact return map (SM~\S3) gives
the survival criterion
\begin{equation}
  \frac{n_1}{n_2}\le\sqrt{1-\gamma},
  \qquad
  p_c(\gamma)=\frac{\sqrt{1-\gamma}}{1+\sqrt{1-\gamma}}.
  \label{eq:pc}
\end{equation}
At $\gamma=0$ this permits only equal clusters, recovering the divisor
principle. For every $\gamma<0$ an unequal window opens, and sufficiently
near-balanced integer splits persist. At $N=7$, $\gamma=-1$, for example,
$4/3<\sqrt2$ and $\approx22\%$ of initial conditions reach the $(4,3)$
attractor. Such cluster states are familiar from pulse-coupled
networks~\cite{Timme2002,Zumdieck2004}; here they are organized by the
arithmetic of $N$.

Physically, convex charging captures the accelerating spike-initiation regime
of quadratic and exponential integrate-and-fire
neurons~\cite{Ermentrout1996,FourcaudTrocme2003,Badel2008}, placing the
critical point at the boundary between leaky (synchronizing) and active
(cluster-forming) dynamics. The fragility is itself the message: the arithmetic
is a critical phenomenon of the charging nonlinearity, not a generic feature of
pulse coupling. Open directions include a closed form for $\Psync(N,\gamma)$ in
the convex interior, the basin geometry of the unequal-cluster attractors, and
whether the critical-point arithmetic recurs in other coalescent systems with
proportional, size-dependent merging.

\end{document}